%% file: main.tex
\documentclass[11pt,a4paper,onecolumn]{article}
\usepackage{marginnote}
\usepackage{graphicx}
\usepackage{subcaption}
\usepackage{xcolor}
\usepackage{authblk,etoolbox}
\usepackage{titlesec}
\usepackage{calc}
\usepackage{tikz}
\usepackage{hyperref}
\hypersetup{colorlinks,breaklinks=true,
            urlcolor=[rgb]{0.0, 0.5, 1.0},
            linkcolor=[rgb]{0.0, 0.5, 1.0}}
\usepackage{caption}
\usepackage{tcolorbox}
\usepackage{amssymb,amsmath}
\usepackage{ifxetex,ifluatex}
\usepackage{seqsplit}
\usepackage{xstring}
\usepackage{longtable}
\usepackage{lscape}
\usepackage[numbers]{natbib}
\usepackage{floatrow}
\usepackage{svg}
\usepackage{float}
\usepackage{nameref}

\usepackage[margin=3cm]{geometry}

\titleformat{\section}
  {\normalfont\sffamily\Large\bfseries}
  {}{0pt}{}
\titleformat{\subsection}
  {\normalfont\sffamily\large\bfseries}
  {}{0pt}{}
\titleformat{\subsubsection}
  {\normalfont\sffamily\bfseries}
  {}{0pt}{}
\titleformat*{\paragraph}
  {\sffamily\normalsize}

\usepackage{hyperref}
\hypersetup{unicode=true,
            pdftitle={DTW-C++: A C++ software for fast Dynamic Time Wrapping Clustering},
            pdfborder={0 0 0},
            breaklinks=true}
\title{Fast dynamic time warping and clustering in C++}

        \author[1]{Volkan Kumtepeli}
          \author[1]{Rebecca Perriment}
          \author[1]{David A. Howey}
    
      \affil[1]{Department of Engineering Science, University of Oxford,
OX1 3PJ, Oxford, UK}
  \date{}

\begin{document}
\maketitle

\section{Abstract}
\label{summary}
We present an approach for computationally efficient dynamic time warping (DTW) and clustering of time-series data. The method frames the dynamic warping of time series datasets as an optimisation problem solved using dynamic programming, and then clusters time series data by solving a second optimisation problem using mixed-integer programming (MIP). There is also an option to use k-medoids clustering for increased speed, when a certificate for global optimality is not essential. The improved efficiency of our approach is due to task-level parallelisation of the clustering alongside DTW. Our approach was tested using the UCR Time Series Archive, and was found to be, on average, 33\% faster than the next fastest option when using the same clustering method. This increases to 64\% faster when considering only larger datasets (with more than 1000 time series). The MIP clustering is most effective on small numbers of longer time series, because the DTW computation is faster than other approaches, but the clustering problem becomes increasingly computationally expensive as the number of time series to be clustered increases.

\section{Introduction}
\label{statement-of-need}
Time series datasets are ubiquitous in science, engineering and many other fields such as economics. Applications range from finding patterns in energy consumption to detecting brain activity in medical applications and discovering patterns in stock prices in the financial industry. Tools for analysing time-series data are widely available, and one such tool is clustering---a form of unsupervised learning that groups datasets into 'similar' subsets, providing useful insights. 

Most time series clustering algorithms depend on dimension reduction or feature extraction techniques to achieve computational efficiency at scale \cite{Aghabozorgi2015Time-seriesReview} but these can introduce bias into the clustering. Distance-based approaches have the significant advantage of directly using the raw data, thus the results are not biased by the feature selection process. However, choosing which distance metric to use is not obvious, and an incorrect choice can lead to illogical clusters. Dynamic time warping \cite{Sakoe1978DynamicRecognition} is a well-known technique for manipulating time series to enable comparisons between datasets, using  local warping (stretching or compressing along the time axis) of the elements within each time series to find an optimal alignment between series. This emphasises the similarity of the shapes of the respective time series, rather than the exact alignment of specific features. Finding similarities in shape is often preferable to finding similarities in time whenever time of occurrence is not relevant to the clustering problem \cite{Aghabozorgi2015Time-seriesReview}. The approach can distinguish similarity in time series when lags or shifts in time occur; these are undetectable if using Euclidean distances \cite{Begum2015AcceleratingStrategy}. This is beneficial even when using time series of the same length and time-frame, such as power load demand time series \cite{Ausmus2020ImprovingWarping}. Finally, a user-defined warping constraint allows flexibility on which time shifts or lags can be defined as `similar' for each clustering problem \cite{Ratanamahatana2005ThreeMining}. The warping constraint uses a `window' to limit which points in one data set can be mapped to another \cite{Sakoe1978DynamicRecognition}. For example, a warping window of 99 means the first data point in one time series can be mapped only up to the hundredth data point in the time series it is being compared to.

Unfortunately, DTW does not scale well in computational speed as the length and number of time series to be compared increases---the computational complexity grows quadratically with the total number of data points. This complexity is a barrier to DTW being widely implemented in time series clustering  \cite{Rajabi2020ASegmentation}. In this paper, we present a novel approach to speed up the computation of DTW distances and the subsequent clustering problem, allowing longer time series and larger datasets to be analysed. We use dynamic programming to solve the DTW problem and then perform clustering of the warped time series, using the pairwise DTW distances, by formulating the clustering problem as a mixed-integer program (MIP). The user must specify the number of clusters required, and the algorithm then finds the optimal clusters, including a centroid for each cluster, where the centroid is the time series within each cluster that minimises the intercluster distance, i.e., the sum of the distances between each time series within the cluster and the respective centroid. The software associated with this paper, \texttt{DTW-C++}, is freely available from \href{https://github.com/Battery-Intelligence-Lab/dtw-cpp}{\texttt{https://github.com/Battery-Intelligence-Lab/dtw-cpp}}. 

While there are other packages available for time series clustering using DTW \cite{Meert2020Dtaidistance, Tavenard2020TslearnData}, \texttt{DTW-C++} offers significant improvements in speed and memory use, especially for larger datasets. As an aside, there are also innovative methods for speeding up DTW by solving approximate versions of the problem. For example, Deriso and Boyd \cite{Deriso2022AWarping} considered DTW as a continuous-time optimal control problem and solved this by discretisation with iterative refinement using regularisation instead of hard band constraints. 

In our approach, speed-up is achieved by task-level parallelisation, allowing multiple pairwise comparisons between time series to be evaluated simultaneously. Additionally, \texttt{DTW-C++} implements more
efficient memory management by solving the DTW problem using only the preceding vector rather than storing the entire warping matrix (see \nameref{mathmatical-background} for details). This means that the complete warping path between each time series is not stored---but this is not required for the clustering process since only the final cost is needed. Reduction in memory use also paves the way for a future GPU implementation of the algorithm  \cite{Schmidt2020CuDTW++:GPUs}. Our approach uses MIP for clustering---this is preferable to other DTW clustering packages that use k-based methods since the iterative nature of the latter means they are susceptible to getting stuck in local optima, whereas MIP provides a certificate for global optimality. However, where a global optimality certificate is not required, \texttt{DTW-C++} also provides the necessary functions to solve the clustering problem iteratively.

\section{Overview of method}
\label{current-dtw-c-functionality}

The current functionality of the software is as follows:
\begin{enumerate}
    \item Load time series data from CSV file(s).
    \item Calculate DTW pairwise distances between time series, using a vector based approach to reduce memory use. There is also the option to use a Sakoe-Chiba band to restrict warping in the DTW distance calculation \cite{Sakoe1978DynamicRecognition}. This speeds up the computation time as well as being a useful constraint for some time series clustering scenarios (e.g., if an event must occur within a certain time window to be considered similar).
    \item Produce a distance matrix containing all pairwise comparisons between each time series in the dataset.
    \item Split all time series into a predefined number of clusters, with a representative centroid time series for each cluster. This can be done using MIP or k-medoids clustering, depending on user choice.
    \item Output the clustering cost, which is the sum of distances between every time series within each cluster and its cluster centroid.
    \item Find the silhouette score and elbow score for the clusters in order to aid the user decision on how many clusters, $k$, to include.
\end{enumerate}

\section{Mathematical background}
\label{mathmatical-background}

\begin{figure}
    \centering
    \begin{subfigure}{.5\textwidth}
        \centering
        \includegraphics[width=.9\linewidth]{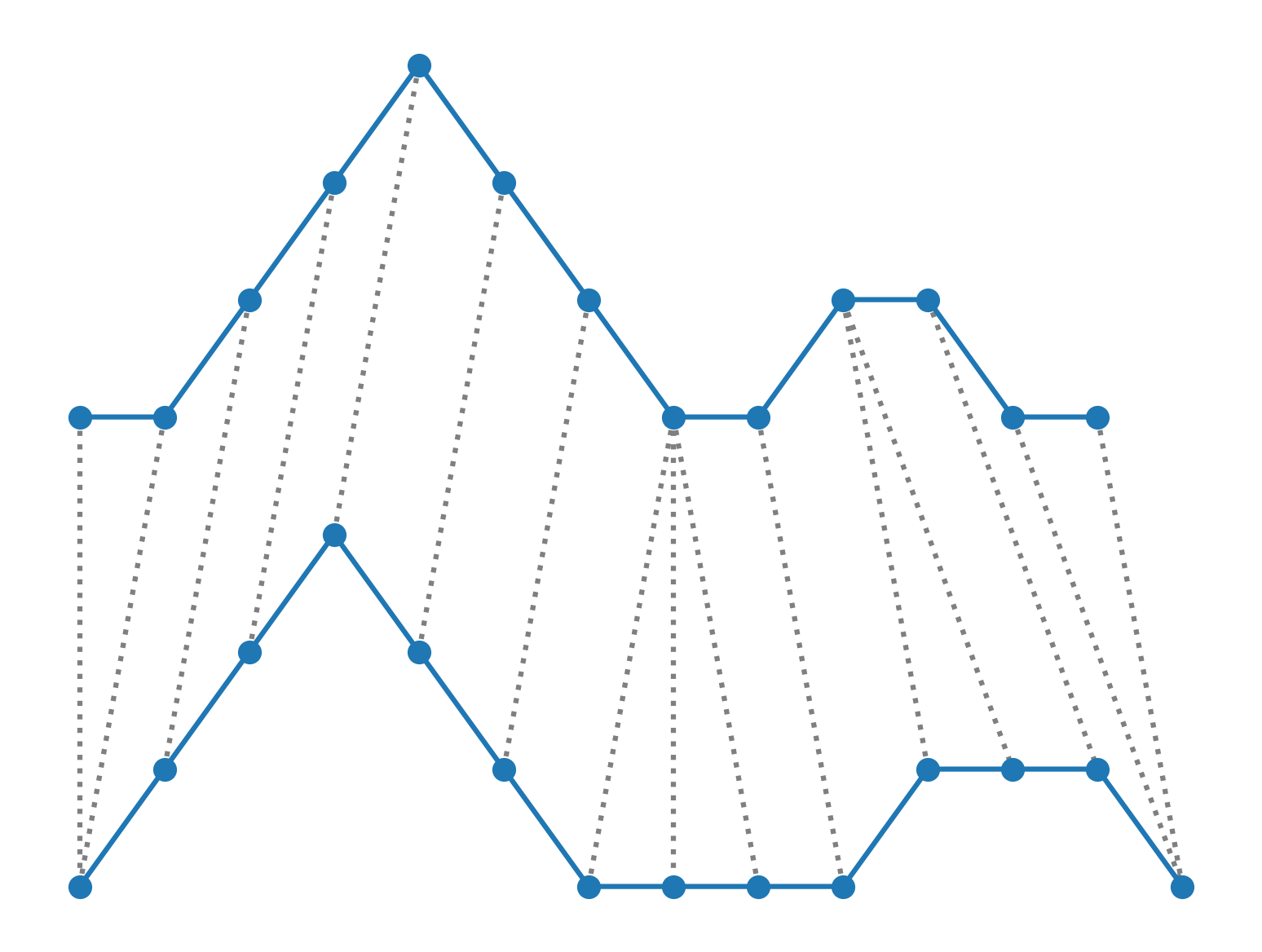}
        \label{fig:sub1}
    \end{subfigure}%
    \begin{subfigure}{.5\textwidth}
      \centering
      \includegraphics[width=.9\linewidth]{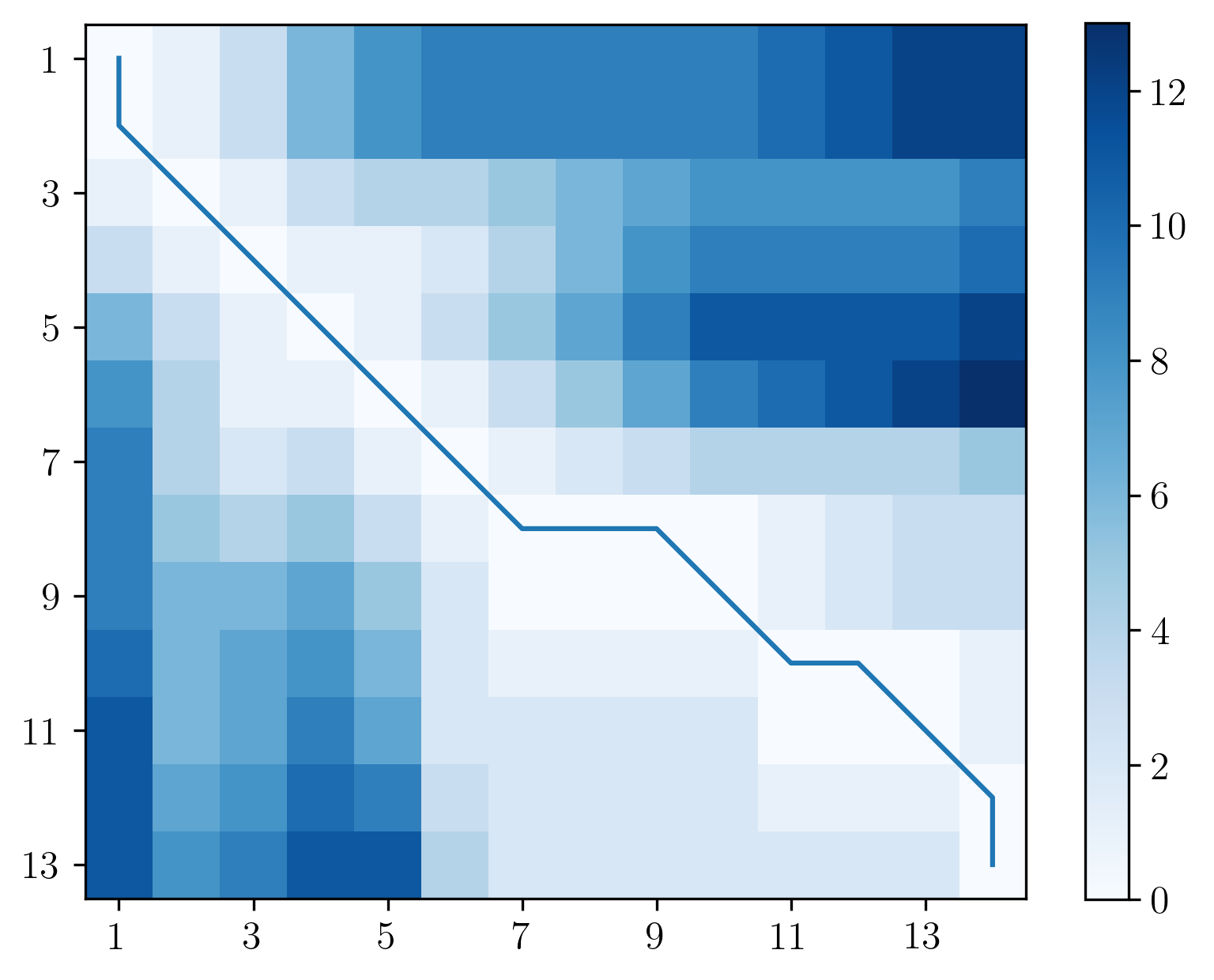}
      \label{fig:sub2}
    \end{subfigure}
    \caption{Two time series with DTW pairwise alignment between each element, showing one-to-many mapping properties of DTW (\emph{left}). Cost matrix $C$ for the two time series, showing the warping path and final DTW cost at $C_{13,12}$ (\emph{right}).}
    \label{fig:warping_signals}
\end{figure}

Consider a time series to be a vector of some arbitrary length. Consider that we have \(p\) such vectors in total, each possibly differing in length. To find a subset of \(k\) clusters within the set of \(p\) vectors using MIP formulation, we must first make $\frac{1}{2} {p \choose 2}$ pairwise comparisons between all vectors within the total set and find the `similarity' between each pair. In this case, the similarity is defined as the DTW distance. Consider two time series \(x\) and \(y\) of differing lengths \(n\) and \(m\) respectively,
\begin{equation}
    x=(x_1, x_2, ..., x_n)
\end{equation}
\begin{equation}
    y=(y_1, y_2, ..., y_m).
\end{equation}
The DTW distance is the sum of the Euclidean distance between each point and its matched point(s) in the other vector, as shown in Fig.\ \ref{fig:warping_signals}. The following constraints must be met: 
\begin{itemize}
    \item The first and last elements of each series must be matched.
    \item Only unidirectional forward movement through relative time is allowed, i.e., if \(x_1\) is mapped to \(y_2\) then \(x_2\) may not be mapped to
    \(y_1\) (monotonicity). 
    \item Each point is mapped to at least one other point, i.e., there are no jumps in time (continuity).
\end{itemize}

Finding the optimal warping arrangement is an optimisation problem that can be solved using dynamic programming, which splits the problem into easier sub-problems and solves them recursively, storing intermediate solutions until the final solution is reached. To understand the memory-efficient method used in \texttt{DTW-C++}, it is useful to first examine the full-cost matrix solution, as follows. For each pairwise comparison, an \(n\) by \(m\) matrix \(C^{n\times m}\) is calculated, where each element represents the cumulative cost between series up to the points \(x_i\) and \(y_j\):
\begin{equation}
    \label{c}
    c_{i,j} = (x_i-y_j)^2+\min\begin{cases}
    c_{i-1,j-1}\\
    c_{i-1,j}\\
    c_{i,j-1}
    \end{cases}
\end{equation}

The final element \(c_{n,m}\) is then the total cost, $C_{x,y}$, which provides the comparison metric between the two series $x$ and $y$. Fig.\ \ref{fig:warping_signals} shows an example of this cost matrix $C$ and the warping path through it. %
For the clustering problem, only this final cost for each pairwise comparison is required; the actual warping path (or mapping of each point in one time series to the other) is superfluous for k-medoids clustering. The memory complexity of the cost matrix $C$ is $\mathcal{O}(nm)$, so as the length of the time series increases, the memory required increases greatly. Therefore, significant reductions in memory can be made by not storing the entire $C$ matrix. When the warping path is not required, only a vector containing the previous row for the current step of the dynamic programming sub-problem is required (i.e., the previous three values $c_{i-1,j-1}$, $c_{i-1,j}$, $c_{i,j-1}$), as indicated in Eq.\ \ref{c}.

\begin{figure}
\centering
\includegraphics[width=1\linewidth]{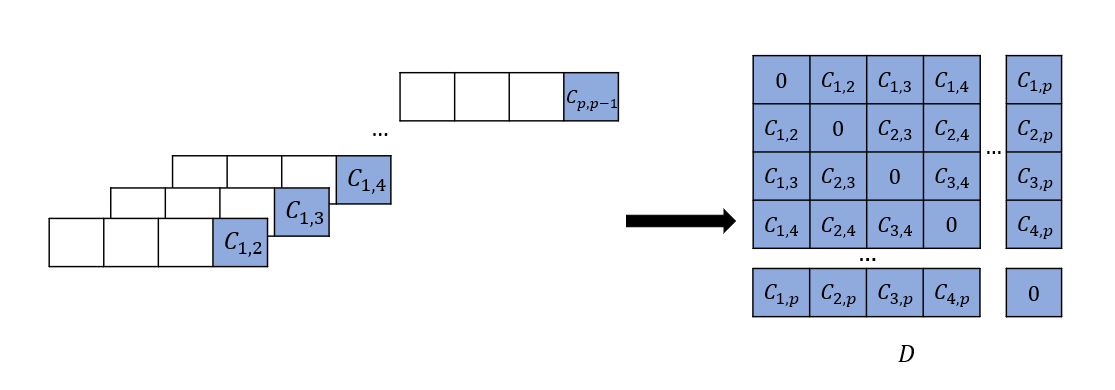}
\caption{The DTW costs of all the pairwise comparisons between time series in the dataset are combined to make a distance matrix $D$.}
\label{fig:c_to_d}
\end{figure}

\begin{figure}
\centering
\includegraphics[width=0.75\linewidth]{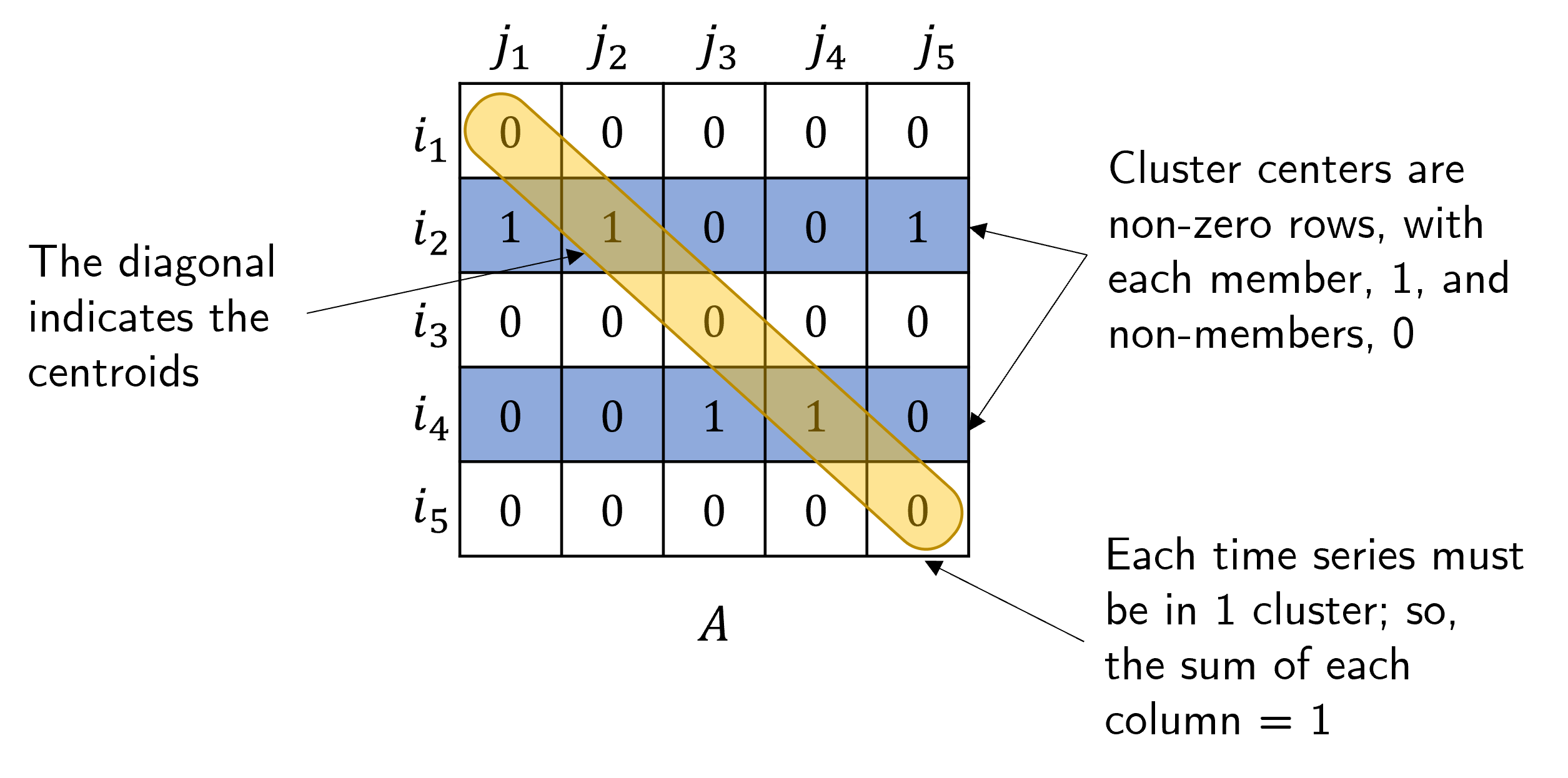}
\caption{Example output from the clustering process, where an entry of 1 indicates that time series $j$ belongs to cluster with centroid $i$}
\label{fig:A_matrix}
\end{figure}

The DTW distance \(C_{x,y}\) is found for each pairwise comparison. As shown in Fig.\ \ref{fig:c_to_d}, pairwise distances are then stored in a separate symmetric matrix, \(D^{p\times p}\), where \(p\) is the total number of time series in the clustering exercise. In other words, the element \(d_{i,j}\) gives the distance between time series \(i\) and \(j\). %
Using this matrix, \(D\), the time series can be split into \(k\) separate clusters with integer programming. The problem formulation begins with a binary square matrix \(A^{p\times p}\), where \(A_{ij}=1\) if time series \(j\) is a member of the \(i\)th cluster centroid, and 0 otherwise, as shown in Fig.\ \ref{fig:A_matrix}. %
As each centroid has to be in its own cluster, non-zero diagonal entries in  $A$ represent centroids. In summary, the following constraints apply: 
\begin{enumerate}
    \item Only \(k\) series can be centroids,
\begin{equation}
    \sum_{i=1}^p A_{ii}=k.
\end{equation}
    \item Each time series must be in one and only one cluster,
\begin{equation}
    \sum_{i=1}^pA_{ij}=1  \quad \forall j \in [1,p].
\end{equation}
\item In any row, there can only be non-zero entries if the corresponding diagonal entry is non-zero, so a time series can only be in a cluster where the row corresponds to a centroid time series,
\begin{equation}
    A_{ij} \le A_{ii} \quad \forall i,j \in [1,p].
\end{equation}
\end{enumerate}
The optimisation problem to solve, subject to the above constraints, is
\begin{equation}
    A^\star = \min_{A} \sum_i \sum_j D_{ij} \times A_{ij}.
\end{equation}

After solving this integer program, the non-zero diagonal entries of \(A\) represent the centroids, and the non-zero elements in the corresponding columns in \(A\) represent the members of that cluster. In the example in Fig.\ \ref{fig:A_matrix}, the clusters are time series 1, \textbf{2}, 5 and 3, \textbf{4} with the bold time series being the centroids.

Finding global optimality can increase the computation time, depending on the number of time series within the dataset and the DTW distances. Therefore, there is also a built-in option to cluster using k-medoids, as used in other packages such as \texttt{DTAIDistance} \cite{Meert2020Dtaidistance}. The k-medoids method is often quicker as it is an iterative approach, however it is subject to getting stuck in local optima. The results in the next section show the timing and memory performance of both MIP clustering and k-medoids clustering using \texttt{DTW-C++} compared to other packages.

\section{Performance comparison and discussion}
\label{comparison}

We compared our approach with two other DTW clustering packages, \texttt{DTAIDistance} \cite{Meert2020Dtaidistance} and \texttt{TSlearn} \cite{Tavenard2020TslearnData}. The datasets used for the comparison are from the UCR Time Series Classification Archive \cite{Dau2018TheArchive}, and consist of 128 time series datasets with up to 16,800 data series of lengths up to 2,844. The full results can be found in Table \ref{tab:big_table} in the Appendix. Benchmarking against  \texttt{TSlearn}  was stopped after the first 22 datasets because the results were consistently over 20 times slower than \texttt{DTW-C++}. Table \ref{tab:small_table} shows the results for datasets downselected to have a number of time series ($N$) greater than 100 and a length of each time series greater than 500 points. This is because \texttt{DTW-C++} is aimed at larger datasets where the speed improvements are more relevant.

\include{summary_table}

\begin{figure}
    \centering
    \includegraphics{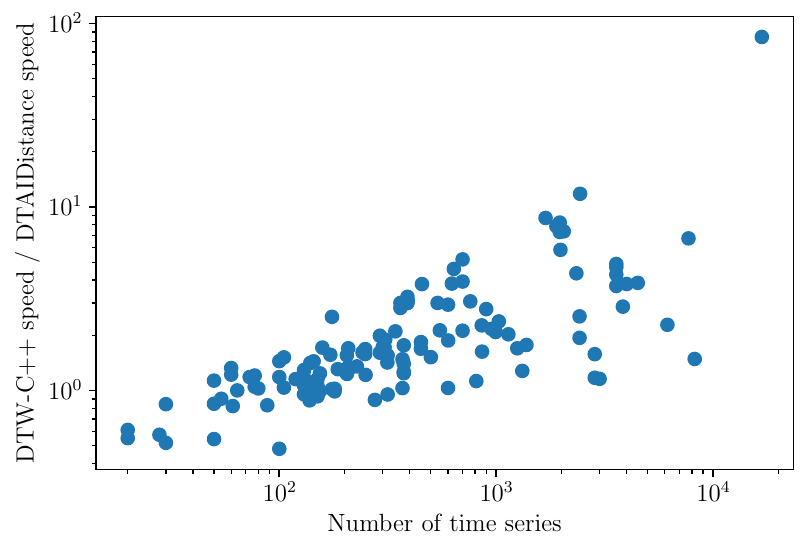}
    \caption{\texttt{DTW-C++} k-medoids  clustering becomes increasingly faster compared to \texttt{DTAIDistance} as the number of time series increases.}
    \label{fig:k_med}
\end{figure}

As can be seen in these results, \texttt{DTW-C++} is the fastest package for 90\% of the datasets, and all 13 datasets where \texttt{DTAIDistance} was faster were cases where the entire clustering process was completed in 1.06 seconds or less. Across the whole collection of datasets, \texttt{DTW-C++} was on average 32\% faster. When looking at larger datasets with $N > 1000$, \texttt{DTW-C++} is on average 65\% faster. In all apart from 2 of the 115 cases where \texttt{DTW-C++} is the fastest, it uses the k-medoids algorithm. This is however to be expected as the latter is an iterative clustering method and therefore does not compute all DTW distances. Fig.\ \ref{fig:k_med} clearly shows the increasing superiority of \texttt{DTW-C++} as the number of time series increases. In this comparison, both algorithms use k-medoids, so the speed improvement is due to faster dynamic time warping. 

\begin{figure}
    \centering
    \includegraphics{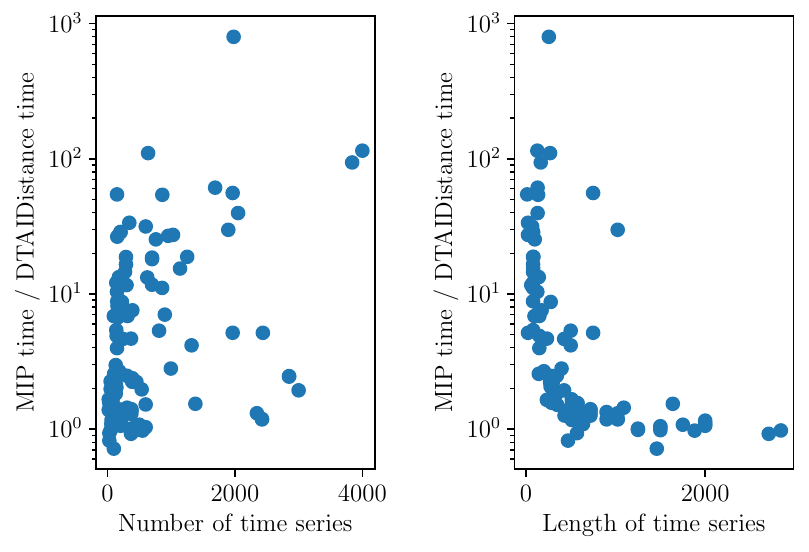}
    \caption{Change in computational time of \texttt{DTW-C++} using MIP DTW clustering compared to \texttt{DTAIDistance} as the number of time series in the datasets to be clustered increases and the length of time series in the datasets increases.}
    \label{fig:mip}
\end{figure}

\texttt{DTW-C++} MIP was on average 16 times slower than \texttt{DTAIDistance} over all samples. Fig.\ \ref{fig:mip} shows that as the number of time series increases, MIP clustering becomes increasingly slower. This is to be expected because the computational complexity of the MIP clustering optimisation increases significantly. However, as the length of the time series increases, the performance of MIP converges to the speed of \texttt{DTAIDistance}, while finding global optimality. This confirms the improved performance of DTW in \texttt{DTW-C++}. Therefore, the MIP approach is recommended for occasions when the time series to be clustered are very long, but the number of time series is smaller. It is also worth noting the length of time series in the UCR Time Series Classification Archive are relatively small compared to many time series datasets, and therefore the performance and relevance of the MIP clustering approach in \texttt{DTW-C++} is understated by these results.

\section{Acknowledgements}
\label{acknowledgements}

We gratefully acknowledge contributions by
\href{https://howey.eng.ox.ac.uk}{Battery Intelligence Lab} members, and thank BBOXX for project funding and access to data. This work was also funded by the UKRI PFER Energy Superhub Oxford demonstrator and the ``Data-driven exploration of the carbon emissions impact of grid energy storage deployment and dispatch'' project (EP/W027321/1).

\bibliographystyle{IEEEtran}
\bibliography{references}

\appendix

\section{Appendix A}
We include here the full benchmarking comparison between \texttt{DTW-C++} (using k-Medoids and MIP), \texttt{DTAIDistance} and \texttt{TSlearn}. As stated in the main text, benchmarking of the latter was discontinued once it was apparent it was significantly slower on all datasets. Additionally, any datasets with a number of time series greater than 4000 were not included for the MIP clustering as the computation time is significantly longer and MIP is not suitable to solve these clustering problems.

\begin{landscape}

\setcounter{table}{0}
\renewcommand{\thetable}{A\arabic{table}}

\scriptsize
\include{benchmarking_table2}
\end{landscape}

\end{document}

%% file: summary_table.tex
\begin{table}[]
\resizebox{\textwidth}{!}{%
\begin{tabular}{l|p{.125\textwidth}p{.125\textwidth}p{.125\textwidth}p{.125\textwidth}p{.125\textwidth}p{.125\textwidth}}
                           & Number of time series    & Length of time series    & DTW-C++ MIP (s) & DTW-C++ k-Medoids (s) & DTAI Distance (s) & Time decrease (\%) \\
\hline
CinCECGTorso               & 1380 & 1639 & 3008.4      & \textbf{1104.2}   & 1955.9       & 44                 \\
Computers                  & 250  & 720  & 16.1        & \textbf{10.5}     & 12.8         & 18                 \\
Earthquakes                & 139  & 512  & 3.2         & \textbf{2.4}      & 2.5          & 3                  \\
EOGHorizontalSignal        & 362  & 1250 & 81.8        & \textbf{27.6}     & 82.9         & 67                 \\
EOGVerticalSignal          & 362  & 1250 & 85.9        & \textbf{30.2}     & 85.2         & 65                 \\
EthanolLevel               & 500  & 1751 & 325.7       & \textbf{198.9}    & 302.3        & 34                 \\
HandOutlines               & 370  & 2709 & 383.7       & \textbf{280.9}    & 415.9        & 32                 \\
Haptics                    & 308  & 1092 & 65.5        & \textbf{24.0}     & 45.5         & 47                 \\
HouseTwenty                & 119  & 2000 & 23.8        & \textbf{19.1}     & 22.0         & 13                 \\
InlineSkate                & 550  & 1882 & 412.4       & \textbf{198.9}    & 423.4        & 53                 \\
InsectEPGRegularTrain      & 249  & 601  & 12.3        & \textbf{5.6}      & 8.9          & 37                 \\
InsectEPGSmallTrain        & 249  & 601  & 11.6        & \textbf{5.3}      & 8.9          & 41                 \\
LargeKitchenAppliances     & 375  & 720  & 44.6        & \textbf{25.6}     & 31.8         & 20                 \\
Mallat                     & 2345 & 1024 & 2948.7      & \textbf{517.0}    & 2251.3       & 77                 \\
MixedShapesRegularTrain    & 2425 & 1024 & 2811.8      & \textbf{1221.9}   & 2367.1       & 48                 \\
MixedShapesSmallTrain      & 2425 & 1024 & 2793.7      & \textbf{934.0}    & 2369.3       & 61                 \\
NonInvasiveFetalECGThorax1 & 1965 & 750  & 52599.0     & \textbf{128.7}    & 941.9        & 86                 \\
NonInvasiveFetalECGThorax2 & 1965 & 750  & 4905.4      & \textbf{115.6}    & 951.0        & 88                 \\
Phoneme                    & 1896 & 1024 & 46549.0     & \textbf{198.4}    & 1560.6       & 87                 \\
PigAirwayPressure          & 208  & 2000 & 84.6        & \textbf{56.7}     & 73.2         & 23                 \\
PigArtPressure             & 208  & 2000 & 78.9        & \textbf{41.8}     & 71.1         & 41                 \\
PigCVP                     & 208  & 2000 & 73.5        & \textbf{51.7}     & 69.5         & 26                 \\
RefrigerationDevices       & 375  & 720  & 36.8        & \textbf{20.3}     & 28.4         & 28                 \\
ScreenType                 & 375  & 720  & 38.6        & \textbf{16.1}     & 28.5         & 43                 \\
SemgHandGenderCh2          & 600  & 1500 & 335.9       & \textbf{315.2}    & 325.4        & 3                  \\
SemgHandMovementCh2        & 450  & 1500 & 177.7       & \textbf{107.2}    & 181.1        & 41                 \\
SemgHandSubjectCh2         & 450  & 1500 & 186.4       & \textbf{96.7}     & 177.6        & 46                 \\
ShapesAll                  & 600  & 512  & 67.5        & \textbf{15.1}     & 44.4         & 66                 \\
SmallKitchenAppliances     & 375  & 720  & 41.7        & \textbf{23.8}     & 30.1         & 21                 \\
StarLightCurves            & 8236 & 1024 & N/A         & \textbf{18551.7}  & 27558.1      & 33                 \\
UWaveGestureLibraryAll     & 3582 & 945  & N/A         & \textbf{1194.6}   & 4436.9       & 73                
\end{tabular}}
\caption{Computational time comparison of \texttt{DTW-C++} using MIP and k-medoids, vs.\ \texttt{DTAIDistance}, and \texttt{TSlearn}, on datasets in the UCR Time Series Classification Archive where $N>100$ and $L>500$.}
\label{tab:small_table}
\end{table}

%% file: benchmarking_table2.tex
\begin{longtable}[c]{p{.30\textwidth} | p{.11\textwidth} p{.11\textwidth} p{.11\textwidth} p{.11\textwidth} p{.11\textwidth} p{.11\textwidth}}
\setcounter{LTchunksize}{100}
\kill

\caption{Speed comparison of \texttt{DTW-C++} using MIP and k-Medoids, DTAIDistance and TSlearn on all datasets in the UCR Time Series Classification Archive.} \\

& Number of time series & Length of time series & DTW-C++ MIP (s) & DTW-C++ k-Medoids (s) & DTAIDistance (s) & TSlearn (s) \\
\hline
\endfirsthead

& Number of time series & Length of time series & DTW-C++ MIP (s) & DTW-C++ k-Medoids (s) & DTAIDistance (s) & TSlearn (s) \\
\hline
\endhead

ACSF1                          & 100                & 1460   & 10.3943     & 10.038            & 14.50846     & 389.8806 \\
Adiac                          & 391                & 176    & 29.304      & 1.26198           & 3.874026     & 172.4454 \\
AllGestureWiimoteX             & 700                & Vary   & 200.949     & 2.08638           & 10.82009     &          \\
AllGestureWiimoteY             & 700                & Vary   & 104.812     & 2.73356           & 5.791508     &          \\
AllGestureWiimoteZ             & 700                & Vary   & 63.5701     & 1.38316           & 5.431429     &          \\
ArrowHead                      & 175                & 251    & 2.27864     & 0.897899          & 0.913173     & 60.84465 \\
Beef                           & 30                 & 470    & 0.148394    & 0.214371          & 0.180589     & 9.44669  \\
BeetleFly                      & 20                 & 512    & 0.107265    & 0.127118          & 0.077477     & 13.51861 \\
BirdChicken                    & 20                 & 512    & 0.121148    & 0.132698          & 0.072962     & 7.080103 \\
BME                            & 150                & 128    & 2.54826     & 0.231731          & 0.245758     & 28.88994 \\
Car                            & 60                 & 577    & 0.764395    & 0.402418          & 0.49119      & 53.99938 \\
CBF                            & 900                & 128    & 52.708      & 2.69667           & 7.495611     & 264.24   \\
Chinatown                      & 343                & 24     & 9.61945     & 0.136138          & 0.286097     & 12.98259 \\
ChlorineConcentration          & 3840               & 166    & 18894.9     & 70.1711           & 201.1307     & 1890.485 \\
CinCECGTorso                   & 1380               & 1639   & 3008.41     & 1104.24           & 1955.915     & 28990.66 \\
Coffee                         & 28                 & 286    & 0.091578    & 0.101924          & 0.058483     & 4.686621 \\
Computers                      & 250                & 720    & 16.1162     & 10.5249           & 12.81184     & 860.0778 \\
CricketX                       & 390                & 300    & 13.4394     & 1.85138           & 6.000062     & 173.9918 \\
CricketY                       & 390                & 300    & 13.0524     & 1.87443           & 5.811014     & 192.1667 \\
CricketZ                       & 390                & 300    & 13.9427     & 1.95588           & 5.861976     & 279.3739 \\
Crop                           & 16800              & 46     &             & 77.5763           & 6563.978     & 9618.363 \\
DiatomSizeReduction            & 306                & 345    & 11.6206     & 2.72367           & 4.685899     & 227.1111 \\
DistalPhalanxOutlineAgeGroup   & 139                & 80     & 0.871482    & 0.114135          & 0.161292     & 5.294581 \\
DistalPhalanxOutlineCorrect    & 276                & 80     & 5.43763     & 0.420276          & 0.373589     & 8.474895 \\
DistalPhalanxTW                & 139                & 80     & 1.70307     & 0.127389          & 0.140715     & 5.276537 \\
DodgerLoopDay                  & 80                 & 288    & 0.408889    & 0.242574          & 0.248893     &          \\
DodgerLoopGame                 & 138                & 288    & 1.55933     & 0.709293          & 0.759482     &          \\
DodgerLoopWeekend              & 138                & 288    & 1.5386      & 0.864304          & 0.763335     &          \\
Earthquakes                    & 139                & 512    & 3.17021     & 2.40538           & 2.475174     &          \\
ECG200                         & 100                & 96     & 0.56719     & 0.171865          & 0.082707     &          \\
ECG5000                        & 4500               & 140    &             & 53.3946           & 206.1784     &          \\
ECGFiveDays                    & 861                & 136    & 366.465     & 4.1557            & 6.770701     &          \\
ElectricDevices                & 7711               & 96     &             & 60.5279           & 408.6165     &          \\
EOGHorizontalSignal            & 362                & 1250   & 81.7745     & 27.6169           & 82.88655     &          \\
EOGVerticalSignal              & 362                & 1250   & 85.8957     & 30.2248           & 85.22367     &          \\
EthanolLevel                   & 500                & 1751   & 325.686     & 198.929           & 302.3411     &          \\
FaceAll                        & 1690               & 131    & 2118.18     & 3.9713            & 34.63922     &          \\
FaceFour                       & 88                 & 350    & 0.668328    & 0.53225           & 0.442447     &          \\
FacesUCR                       & 2050               & 131    & 1882.32     & 6.43057           & 47.43581     &          \\
FiftyWords                     & 455                & 270    & 21.3023     & 2.50822           & 9.540631     &          \\
Fish                           & 175                & 463    & 3.62051     & 1.06903           & 2.695585     &          \\
FordA                          & 1320               & 500    & 703.644     & 132.113           & 168.9301     &          \\
FordB                          & 810                & 500    & 347.998     & 57.843            & 65.1031      &          \\
FreezerRegularTrain            & 2850               & 301    & 735.378     & 190.672           & 300.8929     &          \\
FreezerSmallTrain              & 2850               & 301    & 730.621     & 252.548           & 296.348      &          \\
Fungi                          & 186                & 201    & 1.99303     & 0.569607          & 0.744395     &          \\
GestureMidAirD1                & 130                & Vary   & 1.09427     & 0.313373          & 0.367755     &          \\
GestureMidAirD2                & 130                & Vary   & 0.948828    & 0.282304          & 0.364915     &          \\
GestureMidAirD3                & 130                & Vary   & 0.777829    & 0.338532          & 0.360999     &          \\
GesturePebbleZ1                & 172                & Vary   & 3.99312     & 0.378018          & 0.592169     &          \\
GesturePebbleZ2                & 158                & Vary   & 4.81083     & 0.344109          & 0.589917     &          \\
GunPoint                       & 150                & 150    & 1.35765     & 0.29607           & 0.341601     &          \\
GunPointAgeSpan                & 316                & 150    & 7.86469     & 1.11579           & 1.061542     &          \\
GunPointMaleVersusFemale       & 316                & 150    & 7.95605     & 0.745125          & 1.15842      &          \\
GunPointOldVersusYoung         & 315                & 150    & 7.77761     & 0.773618          & 1.09943      &          \\
Ham                            & 105                & 431    & 1.27085     & 0.972879          & 1.009123     &          \\
HandOutlines                   & 370                & 2709   & 383.677     & 280.885           & 415.8791     &          \\
Haptics                        & 308                & 1092   & 65.493      & 24.0428           & 45.48866     &          \\
Herring                        & 64                 & 512    & 0.623086    & 0.533143          & 0.533707     &          \\
HouseTwenty                    & 119                & 2000   & 23.7889     & 19.1              & 22.04339     &          \\
InlineSkate                    & 550                & 1882   & 412.36      & 198.895           & 423.3659     &          \\
InsectEPGRegularTrain          & 249                & 601    & 12.2515     & 5.62819           & 8.897502     &          \\
InsectEPGSmallTrain            & 249                & 601    & 11.6215     & 5.31629           & 8.943162     &          \\
InsectWingbeatSound            & 1980               & 256    & 93853.4     & 20.0948           & 117.4865     &          \\
ItalyPowerDemand               & 1029               & 24     & 56.7128     & 0.871386          & 2.074309     &          \\
LargeKitchenAppliances         & 375                & 720    & 44.5535     & 25.5697           & 31.76365     &          \\
Lightning2                     & 61                 & 637    & 0.844298    & 0.943195          & 0.77562      &          \\
Lightning7                     & 73                 & 319    & 0.518613    & 0.245382          & 0.290609     &          \\
Mallat                         & 2345               & 1024   & 2948.67     & 516.964           & 2251.27      &          \\
Meat                           & 60                 & 448    & 0.484361    & 0.268107          & 0.355596     &          \\
MedicalImages                  & 760                & 99     & 92.8507     & 1.19697           & 3.664919     &          \\
MelbournePedestrian            & 2439               & 24     & 144.153     & 2.36702           & 27.96836     &          \\
MiddlePhalanxOutlineAgeGroup   & 154                & 80     & 1.3446      & 0.122859          & 0.152365     &          \\
MiddlePhalanxOutlineCorrect    & 291                & 80     & 7.59361     & 0.251781          & 0.404614     &          \\
MiddlePhalanxTW                & 154                & 80     & 4.56761     & 0.173091          & 0.172191     &          \\
MixedShapesRegularTrain        & 2425               & 1024   & 2811.82     & 1221.93           & 2367.125     &          \\
MixedShapesSmallTrain          & 2425               & 1024   & 2793.7      & 934.047           & 2369.322     &          \\
MoteStrain                     & 1252               & 84     & 131.375     & 4.10743           & 6.979077     &          \\
NonInvasiveFetalECGThorax1     & 1965               & 750    & 52599       & 128.662           & 941.9007     &          \\
NonInvasiveFetalECGThorax2     & 1965               & 750    & 4905.43     & 115.551           & 950.9619     &          \\
OliveOil                       & 30                 & 570    & 0.196972    & 0.406894          & 0.211004     &          \\
OSULeaf                        & 242                & 427    & 20.9774     & 2.79601           & 4.513287     &          \\
PhalangesOutlinesCorrect       & 858                & 80     & 39.8342     & 1.58711           & 3.595135     &          \\
Phoneme                        & 1896               & 1024   & 46549       & 198.36            & 1560.56      &          \\
PickupGestureWiimoteZ          & 50                 & Vary   & 0.142592    & 0.085126          & 0.072093     &          \\
PigAirwayPressure              & 208                & 2000   & 84.5969     & 56.658            & 73.22682     &          \\
PigArtPressure                 & 208                & 2000   & 78.9084     & 41.8304           & 71.07334     &          \\
PigCVP                         & 208                & 2000   & 73.4538     & 51.724            & 69.45288     &          \\
PLAID                          & 537                & Vary   & 35.9025     & 6.0825            & 18.25728     &          \\
Plane                          & 105                & 144    & 0.515582    & 0.132734          & 0.201192     &          \\
PowerCons                      & 180                & 144    & 4.85703     & 0.368735          & 0.364681     &          \\
ProximalPhalanxOutlineAgeGroup & 205                & 80     & 6.29439     & 0.141562          & 0.220177     &          \\
ProximalPhalanxOutlineCorrect  & 291                & 80     & 7.70703     & 0.234308          & 0.466041     &          \\
ProximalPhalanxTW              & 205                & 80     & 6.66169     & 0.188157          & 0.231558     &          \\
RefrigerationDevices           & 375                & 720    & 36.8233     & 20.3085           & 28.38962     &          \\
Rock                           & 50                 & 2844   & 8.7356      & 7.90198           & 8.94623      &          \\
ScreenType                     & 375                & 720    & 38.5828     & 16.1376           & 28.46865     &          \\
SemgHandGenderCh2              & 600                & 1500   & 335.918     & 315.171           & 325.3503     &          \\
SemgHandMovementCh2            & 450                & 1500   & 177.672     & 107.214           & 181.079      &          \\
SemgHandSubjectCh2             & 450                & 1500   & 186.388     & 96.7203           & 177.5783     &          \\
ShakeGestureWiimoteZ           & 50                 & Vary   & 0.173276    & 0.141293          & 0.076811     &          \\
ShapeletSim                    & 180                & 500    & 4.83641     & 3.37486           & 3.450108     &          \\
ShapesAll                      & 600                & 512    & 67.5375     & 15.1164           & 44.40931     &          \\
SmallKitchenAppliances         & 375                & 720    & 41.7491     & 23.7893           & 30.08536     &          \\
SmoothSubspace                 & 150                & 15     & 5.48273     & 0.108177          & 0.100557     &          \\
SonyAIBORobotSurface1          & 601                & 70     & 45.182      & 0.763845          & 1.43163      &          \\
SonyAIBORobotSurface2          & 953                & 65     & 83.0127     & 1.42315           & 3.08514      &          \\
StarLightCurves                & 8236               & 1024   &             & 18551.7           & 27558.11     &          \\
Strawberry                     & 370                & 235    & 16.4157     & 3.40908           & 3.515086     &          \\
SwedishLeaf                    & 625                & 128    & 54.7593     & 1.07994           & 4.134455     &          \\
Symbols                        & 995                & 398    & 177.772     & 30.3962           & 63.36334     &          \\
SyntheticControl               & 300                & 60     & 4.76377     & 0.237314          & 0.409787     &          \\
ToeSegmentation1               & 228                & 277    & 15.5281     & 1.31453           & 1.779399     &          \\
ToeSegmentation2               & 130                & 343    & 1.71634     & 0.984133          & 0.938638     &          \\
Trace                          & 100                & 275    & 0.729685    & 0.298243          & 0.353099     &          \\
TwoLeadECG                     & 1139               & 82     & 86.9761     & 2.78576           & 5.642078     &          \\
TwoPatterns                    & 4000               & 128    & 15892.5     & 36.3566           & 138.4134     &          \\
UMD                            & 144                & 150    & 1.29514     & 0.183408          & 0.26445      &          \\
UWaveGestureLibraryAll         & 3582               & 945    &             & 1194.61           & 4436.893     &          \\
UWaveGestureLibraryX           & 3582               & 315    &             & 122.541           & 524.8729     &          \\
UWaveGestureLibraryY           & 3582               & 315    &             & 113.072           & 532.3661     &          \\
UWaveGestureLibraryZ           & 3582               & 315    &             & 107.301           & 525.2477     &          \\
Wafer                          & 6164               & 152    &             & 178.151           & 406.4461     &          \\
Wine                           & 54                 & 234    & 0.215293    & 0.145209          & 0.130907     &          \\
WordSynonyms                   & 638                & 270    & 1525.07     & 3.00847           & 13.84392     &          \\
Worms                          & 77                 & 900    & 2.61596     & 1.6201            & 1.9559       &          \\
WormsTwoClass                  & 77                 & 900    & 2.5413      & 2.05345           & 2.154319     &          \\
Yoga                           & 3000               & 426    & 1221.19     & 544.997           & 631.1096     &          \\

\label{tab:big_table}
\end{longtable}